\documentclass[twocolumn,prl,amsmath,amssymb]{revtex4}
 
\usepackage{graphicx}
\usepackage{dcolumn}
\usepackage{bm}
\usepackage{feynmf} 
 
\newcommand{\un}{{\mathbf{1}}}
\newcommand{\bfr}{{\mathbf{r}}}
\newcommand{\degres}{{\mathrm{deg}}}
\newcommand{\dd}{{\mathrm{d}}}
\newcommand{\ee}{{\mathrm{e}}}
\newcommand{\tr}{{\mathrm{tr}}}

\newcommand{\baru}{u^\dagger}
\newcommand{\bara}{{\bar{\alpha}}}
\newcommand{\bareta}{{\bar{\eta}}}
\newcommand{\barpsi}{{\psi^\dagger}}
\newcommand{\barT}{{T^*}}                                                       
\newcommand{\inter}{{\mathrm{int}}}
\newcommand{\scf}{{\mathrm{SCF}}}

\newcommand{\izx}[1]{{}_{\scriptscriptstyle(#1)}}
 
\begin{document}
 
\preprint{APS/123-QED}
 
\title{Quantum field theory of degenerate systems}
 
\author{Christian Brouder}
\affiliation{%
Laboratoire de Min\'eralogie Cristallographie, CNRS UMR 7590, 
Universit\'es Paris 6 et 7, IPGP, case 115, 4 place Jussieu, 
75252 Paris cedex 05, France.
}%
 
\date{\today}
 
\begin{abstract}
To set up a self-consistent quantum field theory of degenerate
systems, the unperturbed state should be described by
a density matrix instead of a pure state. 
This increases the combinatorial complexity of the many-body
equations. Hopf algebraic techniques are used to deal with
this complexity and show that the Schwinger-Dyson equations are
modified in a non-trivial way.  The hierarchy of Green
functions is derived for degenerate systems, and the
case of a single electron in a two-fold degenerate
orbital is calculated in detail.
\end{abstract}
 
\maketitle
 
The degeneracy of quantum systems is the basis of 
important physical phenomena such as superconductivity,
magnetism, Bose-Einstein condensation, Jahn-Teller distortion, etc. 
Degenerate systems are also numerous. For instance,
all systems containing an odd number of
electrons are degenerate by Kramers' theorem \cite{Kramers30}.
Moreover, degenerate systems are a subject of active
research within the density functional approach
\cite{LevyNagy,Yang,UllrichKohn1,Sahni}.

The quantum field theory of degenerate (or quasidegenerate)
systems has a long history \cite{BH58,Brandow,Kuo,LindgrenMorrison}
and developed recently into the two-time Green function
method \cite{Lebigot,Shabaev} and the
covariant-evolution-operator method \cite{Lindgren0,Lindgren1}.
Still, these theories have
a severe drawback: either they are not self-consistent or
they do not preserve the symmetry of the system.
The antagonism between self-consistency and symmetry can
be seen on a simple example. If the Hamiltonian of a system
is spherically symmetric, the energy eigenstates can be 
classified with angular momentum quantum numbers 
$|LM\rangle$. However, if $L\ge 1$, the system is degenerate
and no state $|LM\rangle$ gives a spherically symmetric
charge density \cite{Lieb}. So, if that charge density
is used to construct a self-consistent potential,
spherical symmetry is broken. The only way to recover
the symmetry is to describe the system with the
density matrix $(2L+1)^{-1} \sum_M |LM\rangle\langle LM|$.
In other words, a self-consistent theory of degenerate
systems must use a density matrix instead of a pure state.

In this paper, we set up a quantum field theory
of degenerate systems based on density matrices.
This is a non trivial extension of the nondegenerate theory
because the usual sum over occupied states must be modified,
and the hierarchy of Green functions acquires additional
terms that 
increase considerably the combinatorial complexity of the
calculations. The problem is solved with two technical
tools: the nonequilibrium quantum field theory and the
Hopf algebra of derivations.

{\sl{The generating function}} --
In nonequilibrium quantum field theory \cite{SchwingerJMP,Chou},
the Green functions can be obtained by the functional derivative
of a generating function $Z_\rho$ with respect to double external 
sources (this is equivalent to Keldysh'
closed-time-path method \cite{Keldysh}). 
In this paper we consider nonrelativistic many-body
theory and the external sources $\eta_+,\eta_-,\bareta_+,\bareta_-$
are fermionic. The density matrix is 
$\hat\rho=\sum_{KL} \rho_{LK} |L\rangle\langle K|$  and
the generating function is 
\begin{eqnarray}
Z_\rho &=& \tr \big(\hat\rho S(\bareta_-,\eta_-)^{-1} 
         S(\bareta_+,\eta_+)\big),
\label{Zrho}
\end{eqnarray}
where 
$S(\bareta_\pm,\eta_\pm) = T \exp\big(-i\int_{-\infty}^\infty H^\inter(t)\dd t
+ i\int \bareta_\pm(x)\psi(x) + \barpsi(x)\eta_\pm(x) \dd x \big)$,
is the S-matrix,
$\psi(x)$ is the electron field operator and 
$ H^\inter(t)$ is the interacting Hamiltonian.
According to the functional formulation of nonequilibrium  QFT \cite{Chou},
the generating function can be written $Z_\rho=\ee^{-iD} Z^0_\rho$,
where $D=D_+-D_-$ and $D_\pm$ is the interaction term
$\int_{-\infty}^\infty H^\inter(t)\dd t$ where the fields are 
replaced by functional derivatives with respect 
to the external sources $\bareta_\pm,\eta_\pm$. Finally, 
$Z^0_\rho$ is the generating function for the noninteracting
system: $Z^0_\rho=\exp[-i\int \boldsymbol{\bareta}(x) 
G^0_0(x,y)\boldsymbol{\eta}(y)\dd x\dd y]
\sum_{KL} \rho_{LK} N^0_{KL}$. In this equation,
$\boldsymbol{\eta}$ is a two-dimensional vector with components
$\eta_+$ and $\eta_-$, $G^0_0(x,y)$ is the 2x2 matrix
\begin{eqnarray*}
\left( \begin{array}{cc}
    -i\langle 0 | T\big(\psi(x)\barpsi(y)\big)|0\rangle
       & -i\langle 0 | \barpsi(y)\psi(x)|0\rangle \\
   i\langle 0 | \psi(x)\barpsi(y)|0\rangle
        & -i\langle 0 |
\barT\big(\psi(x)\barpsi(y)\big)|0\rangle
         \end{array}\right),
\end{eqnarray*}
$\barT$ is the anti-time-ordering operator
and 
$N^0_{KL}=\langle K|{:}\exp\big[ i\int
\bareta_d(x)\psi(x)+\barpsi(x)\eta_d(x)\dd x\big]{:}|L\rangle$,
with $\eta_d=\eta_+-\eta_-$.
Notice that $N^0_{KL}$ is the generating function for the
matrix elements of normal products of field operators.
The proof of these formulas can be found in \cite{Chou}.
The two technical problems that will be solved in this paper 
are the explicit calculation of $N^0_{KL}$ and the 
manipulation of $\ee^{-iD}$ to recover a hierarchy of Green
functions for degenerate systems.

To calculate $N^0_{KL}$, we must define precisely
the states $|K\rangle$ and $|L\rangle$.
The solution of the noninteracting Schr\"odinger
equation provides one-electron orbitals
$u_n(\bfr)$ with energy $\epsilon_n$.
An orbital is called a core orbital if it is filled 
in all states $|K\rangle$ and $|L\rangle$,
otherwise, it is called a valence orbital.
The core orbitals are numbered from 1 to $C$,
the valence orbitals from 1 to $M$.
There are $C$ electrons in the core orbitals
and $N<M$ electrons in the valence orbitals.
For example, in the ion Cr$^{3+}$,
there are $C=18$ core orbitals and
$N=3$ electrons in the $M=10$ orbitals
of the degenerate 3d shell.
The states are generated from the vacuum $|0\rangle$
by the action of creation operators $c^\dagger_n$
for core electrons and $v^\dagger_n$ for valence electrons
as
\begin{eqnarray*}
|K\rangle &=& v^\dagger_{i_N}\dots v^\dagger_{i_1} 
             c^\dagger_C\dots c^\dagger_{1}|0\rangle,\\
|L\rangle &=& v^\dagger_{j_N}\dots v^\dagger_{j_1} 
             c^\dagger_C\dots c^\dagger_{1}|0\rangle,
\end{eqnarray*}
where $i_k$, $j_k$ are valence orbitals
(i.e. integers taken in the set
$\{1,\dots,M\}$) ordered so that
$i_1<\dots<i_N$ and $j_1<\dots<j_N$.
A lengthy calculation yields
\begin{eqnarray*}
N^0_{KL} &=& \prod_{k=1}^C (1+\bara_k\alpha_k)
\\&&\times
\exp\Big(\sum_{n=1}^M \frac{\partial^2}{\partial \alpha_n
      \partial \bara_n}\Big) \bara_{j_1} \alpha_{i_1}
      \dots \bara_{j_N} \alpha_{i_N},
\end{eqnarray*}
with $\bara_n=\int \bareta_d(x)u_n(x)\dd x$
and $\alpha_n=\int \baru_n(x)\eta_d(x)\dd x$,
where $u_n(x)=\ee^{-i\epsilon_n t} u_n(\bfr)$,
$\baru_n(x)=\ee^{i\epsilon_n t} \baru_n(\bfr)$
and $x=(t,\bfr)$. Notice that $\alpha_n$ and
$\bara_n$ are anticommuting (i.e. Grassmann) variables
because the sources are anticommuting variables.

For the following, we shall need the quantity
\begin{eqnarray}
W^0_\rho &=& \log Z^0_\rho =
-i\int \boldsymbol{\bareta}(x) 
G^0_0(x,y)\boldsymbol{\eta}(y)\dd x\dd y
\nonumber\\&&
+ \sum_{k=1}^C \bara_k\alpha_k + \log(\rho(\bara,\alpha)),
\label{W0rho}
\end{eqnarray}
where
\begin{eqnarray*}
\rho(\bara,\alpha)&=& 
\exp(\sum_{n=1}^M \frac{\partial^2}{\partial \alpha_n
      \partial \bara_n})
  \sum_{KL} \rho_{LK}\bara_{j_1} \alpha_{i_1}
      \dots \bara_{j_N} \alpha_{i_N}.
\end{eqnarray*}
To derive (\ref{W0rho}), we used the fact that
$\log(1+\bara_k\alpha_k)=\bara_k\alpha_k$, because
$\bara_k$ and $\alpha_k$ are anticommuting variables.
For the same reason, $\log(\rho(\bara,\alpha))$
is a finite polynomial in $\bara$ and $\alpha$.
The term $-i\int \boldsymbol{\bareta}(x) 
G^0_0(x,y)\boldsymbol{\eta}(y)\dd x\dd y$
is linear in $\eta_\pm$ and $\bareta_\pm$.
The sum over core orbitals is also linear
in $\eta_\pm$ and $\bareta_\pm$.
So we write $\log(\rho(\bara,\alpha))=\rho_l(\bara,\alpha)
+\rho_c(\bara,\alpha)$, where $\rho_l(\bara,\alpha)$
gathers the terms of $\log(\rho(\bara,\alpha))$
which are linear in $\eta_\pm$ and $\bareta_\pm$,
and we obtain
$W^0_\rho=
-i\int \boldsymbol{\bareta}(x) 
G^0_\rho(x,y)\boldsymbol{\eta}(y)\dd x\dd y+\rho_c(\bara,\alpha)$,
where $G^0_\rho(x,y)$ is the sum of all the terms
that are linear in  $\eta_\pm$ and $\bareta_\pm$.

In the standard case of a nondegenerate system \cite{Fetter}, 
the electron-hole transformation shows that the free
Green function $G^0(x,x')$ is a sum over occupied states for $t'>t$
and over empty states for $t<t'$. However, this transformation
is usually justified at the first order only, by showing that
it modifies the free Hamiltonian by a constant \cite{Fetter}.
Equation (\ref{W0rho}) shows that the electron-hole transformation
is valid at all orders. In the degenerate
case, additional terms come from $\rho_l(\bara,\alpha)$.

{\sl{The Hopf algebra of derivations}} --
The hierarchy of Green functions will be obtained through
a manipulation of the exponential of functional derivatives
$\ee^{-iD}$. This manipulation will be made
with the help of the Hopf superalgebra \cite{Majid}
of derivations.
We call derivation a functional derivative with respect
to $\eta_\pm(x)$ or $\bareta_\pm(x)$.
We start from the vector space $V$ generated by the basis
elements $\delta/\delta\eta_\pm(x)$ and 
$\delta/\delta\bareta_\pm(x)$. The exterior Hopf algebra
$\Lambda(V)$ is then a standard mathematical object
\cite{Eisenbud} and we recall here its main properties.
The product is the composition of functional derivatives
(e.g.  the product of $\delta/\delta\eta_+(x)$
and $\delta/\delta\bareta_-(y)$ is
$\delta^2/\delta\eta_+(x)\delta\bareta_-(y)$).
This product is anticommutative. 
If an element $D$ of $\Lambda(V)$ is a product of
$n$ derivations, we say that its degree is
$\degres(D)=n$ and its parity is
$|D|=0$ if $n$ is even and $|D|=1$ if $n$ is odd.
An element $D$ is called even (resp. odd) if 
$|D|=0$ (resp. $|D|=1$).
The product satisfies the identity 
$DD'=(-1)^{|D||D'|} D'D$. Thus, an even element $D$
commutes with all elements of $\Lambda(V)$.
We also introduce a unit $\un$ such that
$\un D= D\un=D$ for all $D\in\Lambda(V)$.

The main operation that we shall use is the coproduct.
It is a linear map $\Delta$ from
$\Lambda(V)$ to $\Lambda(V)\otimes\Lambda(V)$ such that
$\Delta \un=\un\otimes\un$ and
$\Delta \partial=\partial\otimes\un+\un\otimes\partial$
if $\partial$ is a derivation. To define the coproduct
of a term $D$ of degree
greater than 1, we use Sweedler's notation for the
coproduct: $\Delta D =\sum D\izx1\otimes D\izx2$
and the recursive definition
\begin{eqnarray*}
\Delta (D D') &=& \sum (-1)^{|D\izx2||D\izx1'|}
  (D\izx1 D\izx1') \otimes (D\izx2 D\izx2').
\end{eqnarray*}
For example, if $\partial$ and $\partial'$ are derivations,
\begin{eqnarray*}
\Delta (\partial \partial') &=&  \partial \partial'\otimes 1
+  1\otimes\partial \partial'
+ \partial\otimes \partial'
- \partial'\otimes\partial.
\end{eqnarray*}
The convenience of the coproduct stems from the
basic identity 
\begin{eqnarray}
D (fg) &=& \sum (-1)^{|D\izx2||f|} (D\izx1 f)(D\izx2 g),
\label{Dfg}
\end{eqnarray}
where $f$ and $g$ are functions of the external
sources and the fermion fields, $\degres(f)=n$
if $f$ is a product of $n$ sources or fields and
$|f|=\degres(f)$ modulo 2.

The hierarchy of Green functions will be derived
from an identity involving the reduced coproduct
with respect to $D$, denoted by $\Delta'$.
It is defined by 
$\Delta'\un=\un\otimes\un$,
$\Delta'D = \Delta D - \un\otimes D - D\otimes \un$
and then recursively by
\begin{eqnarray*}
\Delta'(D^{n+1}) &=&
\sum (-1)^{|D\izx{1'}||D\izx{2'}^n|}
D\izx{1'}^nD\izx{1'}\otimes D\izx{2'}^nD\izx{2'}.
\end{eqnarray*}
The main property of this reduced coproduct is the
following identity, valid for any even element $D$ of
$\Lambda(V)$.
\begin{eqnarray}
\Delta \ee^{-iD} &=& (\Delta'\ee^{-iD}) (\ee^{-iD}\otimes \ee^{-iD}).
\label{Deltap}
\end{eqnarray}

{\sl{The hierarchy of Green functions}} --
The one-body Green function is 
$G(x,y)=(1/Z_\rho) \partial'\partial Z_\rho|_{\eta=\bareta=0}$,
where $\partial=i\delta/\delta\bareta_\pm(x)$ and
$\partial'=\delta/\delta\eta_\pm(y)$.
The definition (\ref{Zrho}) of $Z_\rho$,
and the fact that $\tr(\hat\rho)=1$
yield $Z_\rho|_{\eta=\bareta=0}=1$.
From $Z_\rho=\ee^{-iD} Z^0_\rho$ and
$Z_\rho=\ee^{W^0_\rho}$ we obtain
$\partial Z_\rho=\ee^{-iD}(\partial Z^0_\rho)
=\ee^{-iD}(\partial W^0_\rho) Z^0_\rho$.
Noticing that $W^0_\rho$ is even, we operate now
with $\partial'$ and obtain
\begin{eqnarray*}
G(x,y) &=& \ee^{-iD}\big((\partial'\partial W^0_\rho)
Z^0_\rho\big)
- \ee^{-iD}\big((\partial W^0_\rho) (\partial' Z^0_\rho)\big),
\end{eqnarray*}
all quantities being evaluated at 
$\boldsymbol{\eta}=\boldsymbol{\bareta}=0$.
Now we apply equations (\ref{Dfg}) and (\ref{Deltap}) and we obtain
\begin{eqnarray*}
G(x,y) &=& \sum(\ee^{-iD})\izx{1'}(\partial'\partial W^1_\rho)
(\ee^{-iD})\izx{2'}Z_\rho
\\&&
- \sum (\ee^{-iD})\izx{1'}(\partial W^1_\rho) 
        (\ee^{-iD})\izx{2'}(\partial' Z_\rho),
\end{eqnarray*}
where $W^1_\rho=\ee^{-iD} W^0_\rho$.
The exponential can be expanded
\begin{eqnarray}
G(x,y)  &=&
\sum_{n=0}^{2M-2} \frac{(-i)^n}{n!}
\sum \big(D\izx{1'}^{n}
\partial'\partial W^1_\rho)
\big(D\izx{2'}^{n} Z_\rho\big)
\nonumber\\&&\hspace*{-20mm}
- \sum_{n=0}^{2M-1} \frac{(-i)^n}{n!}
\sum (-1)^{|D\izx{2'}^{n}|} \big(D\izx{1'}^{n}
\partial W^1_\rho)
\big(D\izx{2'}^{n} \partial'Z_\rho\big).
\label{G(x,y)}
\end{eqnarray}
Equation (\ref{G(x,y)}) is the Schwinger-Dyson
equation for degenerate systems. It is the
main result of the paper.
By further differentiating with respect to
external sources and using (\ref{Dfg}), we obtain the hierarchy
of the $n$-body Green functions.
The number of terms of equation (\ref{G(x,y)}) is finite,
but it can be large when $M$ is large. However, the recursive nature
of the coproduct makes it well suited for symbolic-algebra
packages. 
Notice that equation (\ref{G(x,y)}) has to be used
not only for degenerate systems, but more generally
when the unperturbed state cannot be represented 
by a single Slater determinant.

{\sl{The simplest example}}--
As an illustration, we treat in detail the simplest
case of a single electron in a two-fold degenerate
level (i.e. $N=1$ and $M=2$). This example is important
because it corresponds to Kramers' degeneracy.
The Hamiltonian of a nonrelativistic quantum system is
\begin{eqnarray}
H &=& \int \barpsi(\bfr)(-\frac{\Delta}{2m} + U_n(\bfr))
\psi(\bfr)\dd \bfr \nonumber\\&&+ \frac{e^2}{2} 
\int \barpsi(\bfr) \barpsi(\bfr')
\frac{1}{|\bfr-\bfr'|}\psi(\bfr') \psi(\bfr)  \dd \bfr\dd \bfr',
\label{Hamil}
\end{eqnarray}
where $U_n(\bfr)$ is the nuclear potential.
This Hamiltonian will be split into two parts as
$H=H_0+H_1$, where $H_0$ is the first term of $H$
in (\ref{Hamil}) and $H_1$ its second term.
Thus the interaction is described by the differential
operator $D=D_+-D_-$ with
\begin{eqnarray*}
D_\pm &=&  \frac{e^2}{2} \int \dd t \dd \bfr \dd \bfr'
\frac{1}{|\bfr-\bfr'|}
\\&&
\frac{\delta^4}{\delta\eta_\pm(t,\bfr)\delta\eta_\pm(t,\bfr')
\delta\bareta_\pm(t,\bfr')\delta\bareta_\pm(t,\bfr)}.
\end{eqnarray*}
From this definition we obtain $W^1_\rho=\ee^{-iD}W^0_\rho=W^0_\rho$.

The degenerate levels will be called $n=1$ and
$n=2$. The function $\rho(\bara,\alpha)$ is
$\rho(\bara,\alpha)=1+ (1/2)(\bara_1\alpha_1+\bara_2\alpha_2)$.
Thus $\log(\rho(\bara,\alpha))=(1/2)(\bara_1\alpha_1+\bara_2\alpha_2)
-(1/4)\bara_1\alpha_1\bara_2\alpha_2$. The first term
is added to the free Green function by defining
\begin{eqnarray*}
G^0_\rho(x,y) &=& G^0_0(x,y)
+i\rho_l(x,y)
\left( \begin{array}{cc}
    1 & -1 \\
   -1 & 1 \end{array}\right),
\end{eqnarray*}
with $\rho_l(x,y)= (1/2)(u_1(x) u^\dagger_1(y)+ u_2(x) u^\dagger_2(y))$,
the second term is
$\rho_c(\bara,\alpha)=-(1/4)\bara_1\alpha_1\bara_2\alpha_2$.
Equation (\ref{G(x,y)}) is now rewritten in terms of 
Feynman diagrams:
\begin{fmffile}{dg4}
\setlength{\unitlength}{1mm}
\newcommand{\setval}{\fmfset{wiggly_len}{1.5mm}
\fmfset{arrow_len}{2.5mm}
\fmfset{arrow_ang}{13}\fmfset{dash_len}{1.5mm}\fmfpen{0.25mm}
\fmfset{dot_size}{0.8thick}}
\newcommand{\setgras}{\fmfset{wiggly_len}{1.5mm}
\fmfset{arrow_len}{2.5mm}
\fmfset{arrow_ang}{13}\fmfset{dash_len}{1.5mm}\fmfpen{0.5mm}
\fmfset{dot_size}{0.8thick}}
\newcommand{\scs}{\scriptstyle}

\begin{eqnarray}
 \parbox{15mm}{\begin{center}
  \begin{fmfgraph}(14,5)
  \setval
  \fmfforce{0w,0.5h}{v1}
  \fmfforce{1/2w,0.5h}{v2}
  \fmfforce{1w,0.5h}{v3}
  \fmf{fermion}{v3,v2,v1}
  \fmfdot{v1,v2,v3}
  \fmfv{decor.shape=circle,decor.filled=0.5,
        decor.size=80}{v2}
  \end{fmfgraph}
  \end{center}}
&=&
  \parbox{10mm}{\begin{center}
  \begin{fmfgraph*}(8,3)
  \setval
  \fmfleft{v1}
  \fmfright{v2}
  \fmf{fermion}{v2,v1}
  \fmfdot{v1,v2}
  \end{fmfgraph*}
  \end{center}}
+
 \parbox{20mm}{\begin{center}
  \begin{fmfgraph}(20,5)
  \setval
  \fmfforce{0w,0.5h}{v1}
  \fmfforce{1/4w,0.5h}{v2}
  \fmfforce{2/3w,0.5h}{v3}
  \fmfforce{1w,0.5h}{v4}
  \fmfforce{3/8w,1.5h}{x1}
  \fmf{fermion}{v4,v3,v2,v1}
  \fmf{fermion,right=0.3,tension=0.3}{v3,x1}
  \fmf{fermion,right=0.3,tension=0.3}{x1,v3}
  \fmf{boson}{x1,v2}
  \fmfdot{v1,v2,v4,x1}
  \fmfv{decor.shape=circle,decor.filled=0.5,
        decor.size=80}{v3}
  \end{fmfgraph}
  \end{center}}
+
 \parbox{17mm}{\begin{center}
  \begin{fmfgraph}(15,5)
  \setval
  \fmfforce{0w,0.5h}{v1}
  \fmfforce{1/2w,0.5h}{v2}
  \fmfforce{1w,0.5h}{v3}
  \fmfforce{1/2w,2h}{x1}
  \fmfforce{1/3w,1.25h}{x2}
  \fmfforce{2/3w,1.25h}{x3}
  \fmf{fermion}{v3,v2,v1}
  \fmfdot{v1,v3,x2,x3}
  \fmfv{decor.shape=circle,decor.filled=0,
        decor.size=80}{v2}
  \fmfv{decor.shape=circle,decor.filled=0.5,
        decor.size=80}{x1}
  \fmf{fermion,right=0.3,tension=0.3}{x1,x2,v2}
  \fmf{fermion,right=0.3,tension=0.3}{v2,x3,x1}
  \fmf{boson}{x2,x3}
  \end{fmfgraph}
  \end{center}}
\nonumber\\&&+
 \parbox{17mm}{\begin{center}
  \begin{fmfgraph}(15,5)
  \setval
  \fmfforce{0w,0.5h}{v1}
  \fmfforce{1/2w,0.5h}{v2}
  \fmfforce{1w,0.5h}{v3}
  \fmfforce{1/2w,2.3h}{x1}
  \fmfforce{1/2w,1.1h}{x2}
  \fmfforce{1/2w,1.7h}{x3}
  \fmf{fermion}{v3,v2,v1}
  \fmfdot{v1,v3,x2,x3}
  \fmfv{decor.shape=circle,decor.filled=0,
        decor.size=80}{v2}
  \fmfv{decor.shape=circle,decor.filled=0.5,
        decor.size=80}{x1}
  \fmf{fermion,right=0.9,tension=0.3}{v2,x2}
  \fmf{fermion,right=0.9,tension=0.3}{x2,v2}
  \fmf{fermion,right=0.9,tension=0.3}{x1,x3}
  \fmf{fermion,right=0.9,tension=0.3}{x3,x1}
  \fmf{boson}{x2,x3}
  \end{fmfgraph}
  \end{center}}
+
 \parbox{28mm}{\begin{center}
  \begin{fmfgraph}(26,5)
  \setval
  \fmfforce{0w,0.5h}{v1}
  \fmfforce{1/4w,0.5h}{v2}
  \fmfforce{1/2w,0.5h}{x2}
  \fmfforce{3/4w,0.5h}{x1}
  \fmfforce{1w,0.5h}{v3}
  \fmfforce{2/5w,1.5h}{x3}
  \fmf{fermion}{v3,x1,x2,v2,v1}
  \fmfdot{v1,v3,x2,x3}
  \fmfv{decor.shape=circle,decor.filled=0,
        decor.size=80}{v2}
  \fmfv{decor.shape=circle,decor.filled=0.5,
        decor.size=80}{x1}
  \fmf{fermion,right=0.3,tension=0.3}{v2,x3}
  \fmf{fermion,right=0.3,tension=0.3}{x3,v2}
  \fmf{boson}{x2,x3}
  \end{fmfgraph}
  \end{center}}
\nonumber\\&&
+ \parbox{28mm}{\begin{center}
  \begin{fmfgraph}(26,5)
  \setval
  \fmfforce{0w,0.5h}{p1}
  \fmfforce{1/4w,0.5h}{b1}
  \fmfforce{1/2w,0.0h}{p2}
  \fmfforce{1/2w,0.7h}{p3}
  \fmfforce{1/2w,1.5h}{p4}
  \fmfforce{3/4w,0.5h}{g1}
  \fmfforce{3/4w,1.5h}{p5}
  \fmfforce{1w,0.5h}{p6}
  \fmfdot{p1,p2,p3,p4,p5,p6}
  \fmfv{decor.shape=circle,decor.filled=0,
        decor.size=80}{b1}
  \fmfv{decor.shape=circle,decor.filled=0.5,
        decor.size=80}{g1}
  \fmf{fermion}{p6,g1,p2,b1,p1}
  \fmf{fermion}{g1,p3,b1}
  \fmf{fermion}{b1,p4,g1}
  \fmf{fermion,right=0.5,tension=0.3}{g1,p5}
  \fmf{fermion,right=0.5,tension=0.3}{p5,g1}
  \fmf{boson}{p2,p3}
  \fmf{boson}{p4,p5}
  \end{fmfgraph}
  \end{center}}
+ 
  \parbox{28mm}{\begin{center}
  \begin{fmfgraph}(26,5)
  \setval
  \fmfforce{0w,0.5h}{p1}
  \fmfforce{1/4w,0.5h}{b1}
  \fmfforce{1/2w,0.0h}{p2}
  \fmfforce{1/2w,0.7h}{p3}
  \fmfforce{1/2w,1.5h}{p4}
  \fmfforce{3/4w,0.5h}{g1}
  \fmfforce{3/4w,1.5h}{p5}
  \fmfforce{1w,0.5h}{p6}
  \fmfdot{p1,p2,p3,p4,p5,p6}
  \fmfv{decor.shape=circle,decor.filled=0,
        decor.size=80}{b1}
  \fmfv{decor.shape=circle,decor.filled=0.5,
        decor.size=80}{g1}
  \fmf{fermion}{p6,g1,p2,b1,p1}
  \fmf{fermion}{b1,p3,g1}
  \fmf{fermion}{g1,p4,b1}
  \fmf{fermion,right=0.5,tension=0.3}{g1,p5}
  \fmf{fermion,right=0.5,tension=0.3}{p5,g1}
  \fmf{boson}{p2,p3}
  \fmf{boson}{p4,p5}
  \end{fmfgraph}
  \end{center}}
\nonumber\\&&
+
 \parbox{17mm}{\begin{center}
  \begin{fmfgraph}(15,5)
  \setval
  \fmfforce{0w,0.5h}{p1}
  \fmfforce{1/2w,0.5h}{b}
  \fmfforce{1/4w,1.1h}{p2}
  \fmfforce{1/4w,1.7h}{p3}
  \fmfforce{1/2w,2.3h}{g}
  \fmfforce{3/4w,1.1h}{p4}
  \fmfforce{3/4w,1.7h}{p5}
  \fmfforce{1w,0.5h}{p6}
  \fmfdot{p1,p2,p3,p4,p5,p6}
  \fmfv{decor.shape=circle,decor.filled=0,
        decor.size=80}{b}
  \fmfv{decor.shape=circle,decor.filled=0.5,
        decor.size=80}{g}
  \fmf{fermion}{p6,b,p1}
  \fmf{fermion}{b,p2}
  \fmf{fermion}{p4,b}
  \fmf{fermion,right=0.3,tension=0.3}{p2,g}
  \fmf{fermion,right=0.3,tension=0.3}{g,p4}
  \fmf{fermion,right=0.3,tension=0.3}{g,p3}
  \fmf{fermion,right=0.3,tension=0.3}{p3,g}
  \fmf{fermion,right=0.3,tension=0.3}{g,p5}
  \fmf{fermion,right=0.3,tension=0.3}{p5,g}
  \fmf{boson}{p2,p3}
  \fmf{boson}{p4,p5}
  \end{fmfgraph}
  \end{center}}
+ 
  \parbox{28mm}{\begin{center}
  \begin{fmfgraph}(26,5)
  \setval
  \fmfforce{0w,0.5h}{p1}
  \fmfforce{1/4w,0.5h}{b}
  \fmfforce{1/2w,0.5h}{p2}
  \fmfforce{4/10w,-0.5h}{p3}
  \fmfforce{6/10w,-0.5h}{p4}
  \fmfforce{6/10w,1.5h}{p5}
  \fmfforce{3/4w,0.5h}{g}
  \fmfforce{1w,0.5h}{p6}
  \fmfdot{p1,p2,p3,p4,p5,p6}
  \fmfv{decor.shape=circle,decor.filled=0,
        decor.size=80}{b}
  \fmfv{decor.shape=circle,decor.filled=0.5,
        decor.size=80}{g}
  \fmf{fermion}{p6,g,p2,b,p1}
  \fmf{fermion,right=0.3,tension=0.3}{b,p3}
  \fmf{fermion,right=0.3,tension=0.3}{p3,b}
  \fmf{fermion,right=0.3,tension=0.3}{g,p4}
  \fmf{fermion,right=0.3,tension=0.3}{p4,g}
  \fmf{fermion,right=0.3,tension=0.3}{g,p5}
  \fmf{fermion,right=0.3,tension=0.3}{p5,g}
  \fmf{boson}{p3,p4}
  \fmf{boson}{p2,p5}
  \end{fmfgraph}
  \end{center}}
\nonumber\\&&
+ 
  \parbox{28mm}{\begin{center}
  \begin{fmfgraph}(26,5)
  \setval
  \fmfforce{0w,0.5h}{p1}
  \fmfforce{1/4w,0.5h}{b1}
  \fmfforce{1/2w,0.0h}{p2}
  \fmfforce{1/2w,0.7h}{p3}
  \fmfforce{1/2w,1.5h}{p4}
  \fmfforce{3/4w,0.5h}{g1}
  \fmfforce{3/4w,1.5h}{p5}
  \fmfforce{1/2w,-0.5h}{p6}
  \fmfforce{3/4w,-0.5h}{p7}
  \fmfforce{1w,0.5h}{p8}
  \fmfdot{p1,p2,p3,p4,p5,p6,p7,p8}
  \fmfv{decor.shape=circle,decor.filled=0,
        decor.size=80}{b1}
  \fmfv{decor.shape=circle,decor.filled=0.5,
        decor.size=80}{g1}
  \fmf{fermion}{p8,g1,p4,b1,p1}
  \fmf{fermion,right=0.2,tension=0.3}{g1,p3}
  \fmf{fermion,right=0.2,tension=0.3}{p3,g1}
  \fmf{fermion}{b1,p2,g1}
  \fmf{fermion,right=0.3,tension=0.3}{g1,p5}
  \fmf{fermion,right=0.3,tension=0.3}{p5,g1}
  \fmf{fermion}{g1,p6,b1}
  \fmf{fermion,right=0.3,tension=0.3}{g1,p7}
  \fmf{fermion,right=0.3,tension=0.3}{p7,g1}
  \fmf{boson}{p2,p3}
  \fmf{boson}{p4,p5}
  \fmf{boson}{p6,p7}
  \end{fmfgraph}
  \end{center}}.
\label{feynG}
\end{eqnarray}

In this equation, the interacting 1-body Green function is
$G(x,y)=\parbox{15mm}{\begin{center}
  \begin{fmfgraph}(13,2)
  \setval
  \fmfforce{0w,0.5h}{v1}
  \fmfforce{1/2w,0.5h}{v2}
  \fmfforce{1w,0.5h}{v3}
  \fmf{fermion}{v3,v2,v1}
  \fmfdot{v1,v2,v3}
  \fmfv{decor.shape=circle,decor.filled=0.5,
        decor.size=80}{v2}
  \end{fmfgraph}
  \end{center}}$ 
and the same dot with $2n$ fermion lines is the
interacting $n$-body Green function,
the free Green function is
$G^0_\rho(x,y)=
  \parbox{10mm}{\begin{center}
  \begin{fmfgraph*}(8,1)
  \setval
  \fmfleft{v1}
  \fmfright{v2}
  \fmf{fermion}{v2,v1}
  \fmfdot{v1,v2}
  \end{fmfgraph*}
  \end{center}}$, 
the Coulomb interaction term is 
$\parbox{12mm}{\begin{center}
  \begin{fmfgraph}(10,2)
  \setval
  \fmfforce{0w,1.0h}{v1}
  \fmfforce{0w,0.0h}{v2}
  \fmfforce{1/3w,0.5h}{v3}
  \fmfforce{2/3w,0.5h}{v4}
  \fmfforce{1w,1.0h}{v5}
  \fmfforce{1w,0.0h}{v6}
  \fmfdot{v3,v4}
  \fmf{fermion}{v1,v3}
  \fmf{fermion}{v3,v2}
  \fmf{fermion}{v5,v4}
  \fmf{fermion}{v4,v6}
  \fmf{boson}{v3,v4}
  \end{fmfgraph}
  \end{center}}$,
and
$r(x_1,x_2,y_1,y_2)=
\parbox{12mm}{\begin{center}
  \begin{fmfgraph}(10,2)
  \setval
  \fmfforce{0w,1.0h}{v1}
  \fmfforce{0w,0.0h}{v2}
  \fmfforce{1/2w,0.5h}{b}
  \fmfforce{1w,1.0h}{v5}
  \fmfforce{1w,0.0h}{v6}
  \fmfv{decor.shape=circle,decor.filled=0,
        decor.size=80}{b}
  \fmf{fermion}{v1,b}
  \fmf{fermion}{b,v2}
  \fmf{fermion}{v5,b}
  \fmf{fermion}{b,v6}
  \end{fmfgraph}
  \end{center}}$
describes the correlation induced by
the degeneracy of the system:
\begin{eqnarray*}
r(x_1,x_2,y_1,y_2)&=& -\frac{\delta^4 W^1_\rho}{
\delta\bareta_\pm(x_1)\delta\bareta_\pm(x_2)
\delta\eta_\pm(y_1)\delta\eta_\pm(y_2)}
\\&&\hspace*{-7mm} = (-1/4)(u_1(x_1)u_2(x_2)-u_2(x_1)u_1(x_2))
\\&&\hspace*{-3mm} (\baru_1(y_1)\baru_2(y_2)-\baru_2(y_1)\baru_1(y_2)).
\end{eqnarray*}
The first term of (\ref{feynG})
corresponds to $n=0$ in equation (\ref{G(x,y)}),
the next four terms are for $n=1$, the next four terms
for $n=2$ and the last term for $n=3$. The first term and
the terms where
the initial and final points are connected to the 
white dot (i.e. 3, 4 and 8) come from the first
line of equation (\ref{G(x,y)}),
the terms 2, 5, 6, 7, 9 and 10 come from the second line.
For a nondegenerate system, only the first two terms
of the right hand side survive. So we see that degeneracy
adds a large number of terms to the hierarchy.
It should be noticed that if equation (\ref{feynG}) is operated
by $i\partial/\partial t - h_0$
(where $h_0=-\Delta/2m+U_n$), the first term of the right hand
side gives $\delta(x-x')$, the second term remains and all the
other terms vanish because $(i\partial/\partial t - h_0)u_i(x)=0$.
In other words, the differential form of the hierarchy is
the same for degenerate and nondegenerate systems, but
the integral form (\ref{feynG}) is different.
So the additional terms can be seen as boundary conditions
due to the correlation of the system induced by its degeneracy.
Finally, it should be stressed that this hierarchy is
non perturbative. All Green functions involved in equation (\ref{feynG})
are interacting. However, this hierarchy can be used to derive
a perturbative expansion of the Green function (see, e.g. \cite{Hall}).

To solve equation (\ref{feynG}), the hierarchy must be closed
by using one of the standard approximations such as
the random-phase approximation,
the GW-approximation or the Bethe-Salpeter equation.

{\sl{Self-consistency} }--
The previous equations were not self-consistent. To obtain
self-consistent equations the Hamiltonian $H$ will be split into two parts as
$H=H_0+H_1$, with
$H_0=\int \barpsi(\bfr)(-\frac{\Delta}{2m} + U_n(\bfr)-V(\bfr))
\psi(\bfr)\dd \bfr$ and $H_1=H-H_0$. The potential 
$V(\bfr)$ will be determined self-consistently.
The presence of this external field adds the term $D^v_\pm$ to
$D_\pm$, where
\begin{eqnarray*}
D^v_\pm &=&  \int \dd t \dd \bfr V(\bfr)
\frac{\delta^2}{\delta\eta_\pm(t,\bfr)\delta\bareta_\pm(t,\bfr)},
\end{eqnarray*}
and the differential operator describing the interaction is
now $D^\scf=D_+ + D^v_+ - D_- - D^v_-$. This new interaction
does not modify $W^1_\rho=W^0_\rho$, but it adds the following terms 
to the right-hand side of equation (\ref{G(x,y)}):
\begin{eqnarray}
&&
 \parbox{20mm}{\begin{center}
  \begin{fmfgraph}(20,5)
  \setval
  \fmfforce{0w,0.5h}{p1}
  \fmfforce{1/4w,0.5h}{p2}
  \fmfforce{1/4w,0.5h}{x}
  \fmfforce{2/3w,0.5h}{b}
  \fmfforce{1w,0.5h}{p3}
  \fmf{fermion}{p3,b,p2,p1}
  \fmfdot{p1,p2,p3}
  \fmfv{decor.shape=circle,decor.filled=0.5,
        decor.size=80}{b}
  \fmfv{decor.shape=cross,decor.filled=1,
        decor.size=50}{x}
  \end{fmfgraph}
  \end{center}}
+
 \parbox{17mm}{\begin{center}
  \begin{fmfgraph}(15,5)
  \setval
  \fmfforce{0w,0.5h}{p1}
  \fmfforce{1/2w,0.5h}{b}
  \fmfforce{1/2w,2h}{g}
  \fmfforce{1/4w,1.25h}{x}
  \fmfforce{1/2w,1.25h}{p2}
  \fmfforce{3/4w,1.25h}{p3}
  \fmfforce{1w,0.5h}{p4}
  \fmf{fermion}{p4,b,p1}
  \fmfdot{p1,p2,p3,p4}
  \fmfv{decor.shape=circle,decor.filled=0,
        decor.size=80}{b}
  \fmfv{decor.shape=circle,decor.filled=0.5,
        decor.size=80}{g}
  \fmfv{decor.shape=cross,decor.filled=1,
        decor.size=50}{x}
  \fmf{fermion,right=0.3,tension=0.3}{g,x,b}
  \fmf{fermion,right=0.3,tension=0.3}{b,p3,g}
  \fmf{fermion,right=0.4,tension=0.3}{g,p2}
  \fmf{fermion,right=0.4,tension=0.3}{p2,g}
  \fmf{boson}{p2,p3}
  \end{fmfgraph}
  \end{center}}
+
 \parbox{17mm}{\begin{center}
  \begin{fmfgraph}(15,5)
  \setval
  \fmfforce{0w,0.5h}{p1}
  \fmfforce{1/2w,0.5h}{b}
  \fmfforce{1/2w,2h}{g}
  \fmfforce{1/4w,1.25h}{x}
  \fmfforce{1/2w,1.25h}{p2}
  \fmfforce{3/4w,1.25h}{p3}
  \fmfforce{1w,0.5h}{p4}
  \fmf{fermion}{p4,b,p1}
  \fmfdot{p1,p2,p3,p4}
  \fmfv{decor.shape=circle,decor.filled=0,
        decor.size=80}{b}
  \fmfv{decor.shape=circle,decor.filled=0.5,
        decor.size=80}{g}
  \fmfv{decor.shape=cross,decor.filled=1,
        decor.size=50}{x}
  \fmf{fermion,left=0.3,tension=0.3}{b,x,g}
  \fmf{fermion,left=0.3,tension=0.3}{g,p3,b}
  \fmf{fermion,right=0.4,tension=0.3}{g,p2}
  \fmf{fermion,right=0.4,tension=0.3}{p2,g}
  \fmf{boson}{p2,p3}
  \end{fmfgraph}
  \end{center}}
+
 \parbox{15mm}{\begin{center}
  \begin{fmfgraph}(13,5)
  \setval
  \fmfforce{0w,0.5h}{p1}
  \fmfforce{1/2w,0.5h}{b}
  \fmfforce{1/2w,2h}{g}
  \fmfforce{1/3w,1.25h}{x1}
  \fmfforce{2/3w,1.25h}{x2}
  \fmfforce{1w,0.5h}{p2}
  \fmf{fermion}{p2,b,p1}
  \fmfdot{p1,p2}
  \fmfv{decor.shape=circle,decor.filled=0,
        decor.size=80}{b}
  \fmfv{decor.shape=circle,decor.filled=0.5,
        decor.size=80}{g}
  \fmfv{decor.shape=cross,decor.filled=1,
        decor.size=50}{x1,x2}
  \fmf{fermion,left=0.3,tension=0.3}{b,x1,g}
  \fmf{fermion,left=0.3,tension=0.3}{g,x2,b}
  \end{fmfgraph}
  \end{center}}
\nonumber\\&&
+ 
  \parbox{23.5mm}{\begin{center}
  \begin{fmfgraph}(23,5)
  \setval
  \fmfforce{0w,0.5h}{p1}
  \fmfforce{1/4w,0.5h}{b}
  \fmfforce{4/10w,1.5h}{p2}
  \fmfforce{1/2w,0.5h}{x}
  \fmfforce{6/10w,1.5h}{p3}
  \fmfforce{3/4w,0.5h}{g}
  \fmfforce{1w,0.5h}{p4}
  \fmfdot{p1,p2,p3,p4}
  \fmfv{decor.shape=circle,decor.filled=0,
        decor.size=80}{b}
  \fmfv{decor.shape=circle,decor.filled=0.5,
        decor.size=80}{g}
  \fmfv{decor.shape=cross,decor.filled=1,
        decor.size=50}{x}
  \fmf{fermion}{p4,g,x,b,p1}
  \fmf{fermion,right=0.3,tension=0.3}{b,p2}
  \fmf{fermion,right=0.3,tension=0.3}{p2,b}
  \fmf{fermion,right=0.3,tension=0.3}{g,p3}
  \fmf{fermion,right=0.3,tension=0.3}{p3,g}
  \fmf{boson}{p2,p3}
  \end{fmfgraph}
  \end{center}}
+ 
  \parbox{23.5mm}{\begin{center}
  \begin{fmfgraph}(23,5)
  \setval
  \fmfforce{0w,0.5h}{p1}
  \fmfforce{1/4w,0.5h}{b}
  \fmfforce{4/10w,1.5h}{p2}
  \fmfforce{1/2w,0.5h}{x}
  \fmfforce{6/10w,1.5h}{p3}
  \fmfforce{3/4w,0.5h}{g}
  \fmfforce{1w,0.5h}{p4}
  \fmfdot{p1,p2,p3,p4}
  \fmfv{decor.shape=circle,decor.filled=0,
        decor.size=80}{b}
  \fmfv{decor.shape=circle,decor.filled=0.5,
        decor.size=80}{g}
  \fmfv{decor.shape=cross,decor.filled=1,
        decor.size=50}{x}
  \fmf{fermion}{p4,g,x,b,p1}
  \fmf{fermion}{b,p2,g}
  \fmf{fermion}{g,p3,b}
  \fmf{boson}{p2,p3}
  \end{fmfgraph}
  \end{center}}
+ 
  \parbox{23.5mm}{\begin{center}
  \begin{fmfgraph}(23,5)
  \setval
  \fmfforce{0w,0.5h}{p1}
  \fmfforce{1/4w,0.5h}{b}
  \fmfforce{4/10w,1.5h}{p2}
  \fmfforce{1/2w,0.5h}{x}
  \fmfforce{6/10w,1.5h}{p3}
  \fmfforce{3/4w,0.5h}{g}
  \fmfforce{1w,0.5h}{p4}
  \fmfdot{p1,p2,p3,p4}
  \fmfv{decor.shape=circle,decor.filled=0,
        decor.size=80}{b}
  \fmfv{decor.shape=circle,decor.filled=0.5,
        decor.size=80}{g}
  \fmfv{decor.shape=cross,decor.filled=1,
        decor.size=50}{x}
  \fmf{fermion}{p4,g}
  \fmf{fermion}{b,p1}
  \fmf{fermion}{b,x,g}
  \fmf{fermion}{g,p3,b}
  \fmf{fermion}{g,p2,b}
  \fmf{boson}{p2,p3}
  \end{fmfgraph}
  \end{center}}
\nonumber\\&&
+ 
  \parbox{23.5mm}{\begin{center}
  \begin{fmfgraph}(23,5)
  \setval
  \fmfforce{0w,0.5h}{p1}
  \fmfforce{1/4w,0.5h}{b}
  \fmfforce{4/10w,1.5h}{p2}
  \fmfforce{4/10w,-0.5h}{q2}
  \fmfforce{1/2w,0.5h}{x}
  \fmfforce{3/4w,1.5h}{p3}
  \fmfforce{3/4w,-0.5h}{q3}
  \fmfforce{3/4w,0.5h}{g}
  \fmfforce{1w,0.5h}{p4}
  \fmfdot{p1,p2,q2,p3,q3,p4}
  \fmfv{decor.shape=circle,decor.filled=0,
        decor.size=80}{b}
  \fmfv{decor.shape=circle,decor.filled=0.5,
        decor.size=80}{g}
  \fmfv{decor.shape=cross,decor.filled=1,
        decor.size=50}{x}
  \fmf{fermion}{p4,g,x,b,p1}
  \fmf{fermion}{g,p2,b}
  \fmf{fermion}{b,q2,g}
  \fmf{fermion,right=0.3,tension=0.3}{g,p3}
  \fmf{fermion,right=0.3,tension=0.3}{p3,g}
  \fmf{fermion,right=0.3,tension=0.3}{g,q3}
  \fmf{fermion,right=0.3,tension=0.3}{q3,g}
  \fmf{boson}{p2,p3}
  \fmf{boson}{q2,q3}
  \end{fmfgraph}
  \end{center}}
+ 
  \parbox{23.5mm}{\begin{center}
  \begin{fmfgraph}(23,5)
  \setval
  \fmfforce{0w,0.5h}{p1}
  \fmfforce{1/4w,0.5h}{b}
  \fmfforce{4/10w,1.5h}{p2}
  \fmfforce{4/10w,-0.5h}{q2}
  \fmfforce{1/2w,0.5h}{x}
  \fmfforce{3/4w,1.5h}{p3}
  \fmfforce{3/4w,-0.5h}{q3}
  \fmfforce{3/4w,0.5h}{g}
  \fmfforce{1w,0.5h}{p4}
  \fmfdot{p1,p2,q2,p3,q3,p4}
  \fmfv{decor.shape=circle,decor.filled=0,
        decor.size=80}{b}
  \fmfv{decor.shape=circle,decor.filled=0.5,
        decor.size=80}{g}
  \fmfv{decor.shape=cross,decor.filled=1,
        decor.size=50}{x}
  \fmf{fermion}{p4,g}
  \fmf{fermion}{b,p1}
  \fmf{fermion}{b,x,g}
  \fmf{fermion}{g,p2,b}
  \fmf{fermion}{g,q2,b}
  \fmf{fermion,right=0.3,tension=0.3}{g,p3}
  \fmf{fermion,right=0.3,tension=0.3}{p3,g}
  \fmf{fermion,right=0.3,tension=0.3}{g,q3}
  \fmf{fermion,right=0.3,tension=0.3}{q3,g}
  \fmf{boson}{p2,p3}
  \fmf{boson}{q2,q3}
  \end{fmfgraph}
  \end{center}}
+ 
  \parbox{23.5mm}{\begin{center}
  \begin{fmfgraph}(23,5)
  \setval
  \fmfforce{0w,0.5h}{p1}
  \fmfforce{1/4w,0.5h}{b}
  \fmfforce{4/10w,1.5h}{p2}
  \fmfforce{1/2w,0.5h}{x1}
  \fmfforce{1/2w,0h}{x2}
  \fmfforce{3/4w,1.5h}{p3}
  \fmfforce{3/4w,0.5h}{g}
  \fmfforce{1w,0.5h}{p4}
  \fmfdot{p1,p2,p3,p4}
  \fmfv{decor.shape=circle,decor.filled=0,
        decor.size=80}{b}
  \fmfv{decor.shape=circle,decor.filled=0.5,
        decor.size=80}{g}
  \fmfv{decor.shape=cross,decor.filled=1,
        decor.size=50}{x1,x2}
  \fmf{fermion}{p4,g}
  \fmf{fermion}{b,p1}
  \fmf{fermion}{b,x1,g}
  \fmf{fermion}{g,p2,b}
  \fmf{fermion,left=0.2,tension=0.3}{g,x2,b}
  \fmf{fermion,right=0.3,tension=0.3}{g,p3}
  \fmf{fermion,right=0.3,tension=0.3}{p3,g}
  \fmf{boson}{p2,p3}
  \end{fmfgraph}
  \end{center}}
\nonumber\\&&
+ 
  \parbox{23.5mm}{\begin{center}
  \begin{fmfgraph}(23,5)
  \setval
  \fmfforce{0w,0.5h}{p1}
  \fmfforce{1/4w,0.5h}{b}
  \fmfforce{4/10w,1.5h}{p2}
  \fmfforce{1/2w,0.5h}{x1}
  \fmfforce{1/2w,0h}{x2}
  \fmfforce{3/4w,1.5h}{p3}
  \fmfforce{3/4w,0.5h}{g}
  \fmfforce{1w,0.5h}{p4}
  \fmfdot{p1,p2,p3,p4}
  \fmfv{decor.shape=circle,decor.filled=0,
        decor.size=80}{b}
  \fmfv{decor.shape=circle,decor.filled=0.5,
        decor.size=80}{g}
  \fmfv{decor.shape=cross,decor.filled=1,
        decor.size=50}{x1,x2}
  \fmf{fermion}{p4,g}
  \fmf{fermion}{b,p1}
  \fmf{fermion}{g,x1,b}
  \fmf{fermion}{b,p2,g}
  \fmf{fermion,left=0.2,tension=0.3}{g,x2,b}
  \fmf{fermion,right=0.3,tension=0.3}{g,p3}
  \fmf{fermion,right=0.3,tension=0.3}{p3,g}
  \fmf{boson}{p2,p3}
  \end{fmfgraph}
  \end{center}}
+ 
  \parbox{25mm}{\begin{center}
  \begin{fmfgraph}(22,5)
  \setval
  \fmfforce{0w,0.5h}{p1}
  \fmfforce{1/4w,0.5h}{b}
  \fmfforce{1/2w,0.5h}{x1}
  \fmfforce{1/2w,-0.0h}{x2}
  \fmfforce{1/2w,1.0h}{x3}
  \fmfforce{3/4w,0.5h}{g}
  \fmfforce{1w,0.5h}{p4}
  \fmfdot{p1,p4}
  \fmfv{decor.shape=circle,decor.filled=0,
        decor.size=80}{b}
  \fmfv{decor.shape=circle,decor.filled=0.5,
        decor.size=80}{g}
  \fmfv{decor.shape=cross,decor.filled=1,
        decor.size=50}{x1,x2,x3}
  \fmf{fermion}{p4,g}
  \fmf{fermion}{b,p1}
  \fmf{fermion}{b,x1,g}
  \fmf{fermion,left=0.2,tension=0.3}{g,x2,b}
  \fmf{fermion,right=0.2,tension=0.3}{g,x3,b}.
  \end{fmfgraph}
  \end{center}}
\end{eqnarray}

The condition of self-consistency is now that the 
potential $V(\bfr)$ is generated by the
electron charge density of the system or
$\parbox{3mm}{\begin{center}
  \begin{fmfgraph}(2,2)
  \setval
  \fmfforce{0.5w,0.5h}{x}
  \fmfv{decor.shape=cross,decor.filled=1,
        decor.size=50}{x}
  \end{fmfgraph}
  \end{center}}
= -
  \parbox{10mm}{\begin{center}
  \begin{fmfgraph}(8,2)
  \setval
  \fmfforce{0w,0.4h}{p1}
  \fmfforce{1/2w,0.4h}{p2}
  \fmfforce{1w,0.4h}{g}
  \fmfdot{p1,p2}
  \fmf{boson}{p1,p2}
  \fmfv{decor.shape=circle,decor.filled=0.5,
        decor.size=80}{g}
  \fmf{fermion,left=1,tension=0.3}{g,p2}
  \fmf{fermion,left=1,tension=0.3}{p2,g}
  \end{fmfgraph}
  \end{center}}$
\end{fmffile}  
in terms of Feynman diagrams.

In this paper we showed how to calculate the Green function
of a degenerate system, and we derived the corresponding
hierarchy of Green functions.
An advantage of our approach is that 
it preserves the symmetry of 
the Hamiltonian if the closure of the
hierarchy is done properly. Recall that
this is not possible if we use states
instead of density matrices.

Equation (\ref{G(x,y)}) is written in terms of 
unconnected Green functions. By defining a second
type of reduced coproduct, it is possible to write
the hierarchy of connected Green functions.
In the near future, several developments of the
present formalism will be presented.
Firstly, if we add photon fields, the same formalism
can be used to derive the Schwinger-Dyson equations for the 
Green functions of a degenerate systems in quantum
electrodynamics; this will be applied to the
nonperturbative calculation of a degenerate 
atomic system. Secondly, in the derivation of
equation (\ref{G(x,y)}), no special form was
assumed for the density matrix $\hat\rho$. 
So the present formalism can be used also for
out-of-equilibrium quantum field theory. In
particular, we can calculate the energy of the interacting
system corresponding to a general density matrix
$\hat\rho$.  If we minimize
the total energy with respect to $\hat\rho$, we obtain a 
set of equations
that unify the Green function and diagonalization methods
of many-body theory.

\begin{acknowledgments}
I gratefully acknowledge the fruitful comments
by Matteo Calandra, Philippe Sainctavit
and Francesco Mauri.
\end{acknowledgments}


%
\end{document}